\documentstyle[12pt]{article}
\newcommand{\sect}[1]{\setcounter{equation}{0}\section{#1}}

\begin{flushright}
UT-KOMABA/94/13\\
June 1994
\end{flushright}
\begin{document}
\topmargin 0pt
\oddsidemargin 5mm
\headheight 0pt
\topskip 9mm

\begin{center}
{{\Large\bf Renormalized expansions for matrix models}
\footnote{submitted to Prog. Theor. Phys.( invited paper )}}
\end{center}
\vskip 10mm
\begin{center}
{\sl Shinobu HIKAMI}
\end{center}
\vskip 10mm
\begin{center}
Department of Pure and Applied Sciences,\\
University of Tokyo,\\
Meguro-ku, Komaba 3-8-1, Tokyo 153
\end{center}
\vspace{24pt}
\begin{abstract}

   Matrix models of 2d quantum gravity coupled to matter field are
investigated by the renormalized perturbational method, in which
the matrix model Hamiltonian is represented by the equivalent
vector model.  By the saddle point method, the renormalization
group $\beta$-function is obtained in the successive
approximation.
\end{abstract}
%\end{titlepage}

\newpage
\addtolength{\baselineskip}{0.20\baselineskip}

\sect{Introduction}

\par
  The large $N$ limit of matrix model has been studied
in many fields, including 2d quantum gravity, mesoscopic
fluctuation and quantum chromodynamics.
It is known that several matrix models are solvable in the large
$N$ limit.$^{1),2)}$  When the space dimensionality or
the central charge $c$ increases, the matrix model becomes
difficult to solve analytically.  There is a $c=1$ barrier.$^{3)}$
Only perturbational analysis$^{4)\sim 6)}$
or the numerical simulations$^{7)}$
have been investigated for the case $c>1$.

  In this paper, we consider the matrix model by the saddle
point method and we formulate the renormalized expansion for the
matrix model.  This renormalized expansion is based upon the
rewriting the matrix model Hamiltonian by the equivalent $N^2$-vector
model Hamiltonian.  Using the successive approximation for this
Hamiltonian, we derive the ciritical point and the renormalization
group $\beta$-function.$^{8),9)}$  We show that this method is
effective for several matrix models.  The task of deriving the
perturbational series is reduced and the renormalization group
analysis becomes possible.

  This paper is organized as follows.  The section 2 deals with
the one matrix model.  Following the rule of obtaining the
equivalent vector model Hamiltonian, we evaluate the renormalized
expansion for the one matrix model.  We obtain the exact
renormalization group $\beta$-function.  Also we investigate
the phase transition for the negative coefficient of
${\rm tr}M^2$.$^{10)}$  Our formulation is very effective for this
nonperturbative phenomenon.  In section 3, we analyze the critical
point by the successive approximation and discuss the value of the
string susceptibility by the scaling relation.  In section 4,
we consider $d=1$ vector model by transforming the Hamiltonian
to the equivalent $d=1$ vector model.  In section 5, a gauged matrix
model is considered as a model of $c=1$ case.  This model becomes
a good example of our method.  In section 6, two-matrix model is
formulated in this renormalized expansion.  In section 7,
we consider general case of $n$-Ising model on a random surface,
which is represented by the $2^n$-matrix model.  In section 8,
we discuss the renormalization group $\beta$-function in more
details.

%Section 2
\sect{One matrix model}

\par
  As a two dimensional quantum gravity or random surface, one
matrix model has been studied.  In the large $N$ limit, this
model is solvable by Hilbert-Riemann integral equation.$^{1)}$
The Hamiltonian for this one matrix model is given by
%(2.1)
\begin{equation}
  H = \frac{1}{2}{\rm tr} M^2 + \frac{g}{N}{\rm tr} M^4
\end{equation}

\noindent
where $M$ is $N \times N$ Hermitian matrix.  The free energy $F$ is
obtained
%(2.2)
\begin{equation}
  F = -\frac{1}{N^2} \ln Z
\end{equation}
%
%(2.3)
\begin{equation}
  Z = \int {\rm dM} e^{-H} .
\end{equation}
Since the terms of Hamiltonian are diagonalized by the unitary
matrix, the partition function is expressed by the eigenvalues
of the Hermitian matrix $M$.  Although this eigenvalue
representation is standard method, we employ other representation,
which is closely related to the diagrammatic expansion.  As we
will discuss later, eigenvalue representation does not work for
the coupled matrix models.

The perturbational series of the free energy $F$ in the large $N$
limit becomes
%
%(2.4)
\begin{equation}
\frac{F}{N^2} =\frac{1}{2}+2g-18g^2 +288g^3 -6048g^4 + \cdots
\end{equation}
\noindent
This perturbational series is obtained by counting the Feynman
diagrams as Fig.1.
\vspace{6pt}
\begin{center}
Fig.1
\end{center}
\vspace{6pt}

As an equivalent matrix model to the original model of (2.1), we
consider the following matrix model,
%
%(2.5)
\begin{equation}
H_{eff} = \frac{1}{2}{\rm tr}M^2 +\sum_{K=1}^{\infty}
\frac{c_K g^K}{N^{4K-2}}({\rm tr}M^2 )^{2K}.
\end{equation}
\noindent
This model gives the same free energy of the one matrix model in the
large $N$-limit with appropriate coefficients $c_K$.
Since it is written by polynomial of ${\rm tr}M^2$,
this model is essentially $N^2$-vector model with infinite higher
orders.  This effective Hamiltonian is expressed
by the $N^2$ dimensional
vector field $\bf r$, and the integral measure of the partition function is
$r^{N^2 -1}dr$.  In the large $N$ limit, then we have
%
%(2.6)
\begin{equation}
Z = \int dx \, x^{\frac{N^2}{2}} e^{-H_{eff}}
\end{equation}
%
%(2.7)
\begin{equation}
H_{eff} = \frac{1}{2}x + \sum_{K=1}^{\infty}\frac{c_K g^K}{N^{4K-2}}
x^{2K}
\end{equation}
\noindent
where we have
%
%(2.8)
\begin{equation}
x = r^2 = {\rm tr} M^2 .
\end{equation}
By replacing $x \rightarrow N^2 x$, we obtain the free energy
by the saddle point method,
%
%(2.9)
\begin{equation}
F = \frac{1}{2}x + \sum_{K=1}^{\infty}c_K g^K x^{2K}
  - \frac{1}{2}\ln x
\end{equation}
\noindent
and the saddle point equation becomes
%
%(2.10)
\begin{equation}
x\frac{\partial F}{\partial x} = \frac{1}{2}x
+ \sum_{K=1}^{\infty}c_K (2K)g^K x^{2K}-\frac{1}{2}=0.
\end{equation}
\noindent
We denote $F/N^2$ by $F$.
The coefficient $c_K$ is the weight of the irreducible diagrams
as shown in Fig.2 in the original one matrix model.
\vspace{6pt}
\begin{center}
Fig. 2
\end{center}
\vspace{6pt}
\noindent
For the lower order coefficients,
we have $c_1 =2$, $c_2 =-2$, $c_3 =\frac{32}{3}$, $c_4 =-96$ and
$c_5 =1126.4$.  The effective Hamiltonian is also obtained by the
integration of the angular coordinate keeping the radial part from
the original matrix model.

  In the diagrammatic analysis, this rewriting the Hamiltonian
as a form of $N^2$-vector model means
the renormalization of the propagator.  The field variable $x$
is now the fully renormalized propagator and the self-consistent
equation of the renormalization is the saddle point equation
 (2.10).  The terms of the effective Hamiltonian are the
irreducible diagrams written by $x$.  This remarkable property
is useful for the numerical analysis.  We define $y$ by
%
%(2.11)
\begin{equation}
y =  \frac{\partial F}{\partial g}
\end{equation}
\noindent
and it is easy to see the following identity from the saddle
point equation (2.10)
%
%(2.12)
\begin{equation}
x = 1 - 4gy
\end{equation}
\noindent
By the definition of $y$, from the series expression of (2.9)
we have the following equation
%
%(2.13)
\begin{eqnarray}
y & = & \sum K c_K g^{K-1}x^{2K}  \nonumber \\
  & = & 2x^2 -4gx^4 + 32g^2 x^6 -384g^3 x^8 + 5632g^4x^{10}+\cdots
\end{eqnarray}
\noindent
and by the iteration for small $g$, we express $x^2$ by the power
series of $y$ as
%
%(2.14)
\begin{equation}
x^2 = \frac{1}{2}y + \frac{1}{2}gy^2 -g^2y^3 +\frac{9}{2}g^3y^4
-27g^4y^5 + \cdots .
\end{equation}
{}From (2.12) and (2.14), the value of $y$ is determined.
For one matrix model, the
series of (2.14) is expressed by a closed form.  The ratio
$R_K =c_K /c_{K-1}$ of the
successive coefficients of (2.14) is exactly
%(2.15)
\begin{equation}
R_K = -12 + 30/K \; \;\; \; (K \geq 3) .
\end{equation}
\noindent
Then we have from (2.14)
%
%(2.16)
\begin{equation}
gx^2 = \frac{1}{108}\left[ (1+12gy)^{\frac{3}{2}}-1\right]
+\frac{gy}{3}
\end{equation}
\noindent
{}From (2.15) and (2.12), we find the critical point, where
$gy=-1/12$, and $x=4/3$,
%
%(2.17)
\begin{equation}
g_c = -\frac{1}{48}
\end{equation}
\noindent
The quantities $x$ and $x^2$ and also
$y=(\frac{\partial F}{\partial g})$
has the same singularity near $g_c$ as
%
%(2.18)
\begin{equation}
  y \sim (g - g_c )^{1-\gamma_{st}}
\end{equation}
\noindent
with $\gamma_{st} = -1/2$.

The two equations (2.12) and (2.16) determine the singular
behavior of the one-matrix model, i.e. pure 2d quantum gravity.

The advantage of this saddle point formulation is that we obtain
the string susceptibility $\gamma_{st}$ of (2.18) by the
renormalized series expansion of (2.13) or (2.14).  Compared to
the previous unrenormalized series expansion in the power of $g$,
the number of Feynman diagrams are considerably reduced.

We now discuss the renormalization group $\beta$-function.  As
observed in the previous paper,$^{4)}$ the perturbational series
of the free energy in the power of $g$ has a simple recursive
relation.  We find that
%
%(2.19)
\begin{equation}
  F = \sum c_K g^K
\end{equation}
%(2.20)
\begin{equation}
  R_K = \frac{c_K}{c_{K-1}}
      = (-48)\frac{(1-\frac{1}{K})(1-\frac{1}{2K})}{(1+\frac{2}{K})}
\end{equation}
Noting that
%
%(2.21)
\begin{equation}
  g\frac{\partial F}{\partial g} = \sum c_K K g^K
\end{equation}
%
%(2.22)
\begin{equation}
g\frac{\partial F}{\partial g}+g^2 \frac{\partial^2 F}{\partial g^2}
= \sum c_K K^2 g^K
\end{equation}
\noindent
we obtain from (2.20),
%(2.23)
\begin{equation}
3g(1+24g)\biggl ( \frac{\partial F}{\partial g} \biggr )
= -g^2 (1+48g)\biggl ( \frac{\partial^2 F}{\partial g^2} \biggr )+6g.
\end{equation}
\noindent
The last term is added to be consistent with the first coefficient
of (2.4).

The equation (2.23) is
%(2.24)
\begin{equation}
  y = \beta (g)\biggl ( \frac{\partial y}{\partial g}\biggr )
+ \frac{2}{1+24g}
\end{equation}
\noindent
and
%(2.25)
\begin{equation}
  \beta (g) = -\frac{g}{3}\frac{(1+48g)}{(1+24g)}.
\end{equation}
\noindent
The renormalization group equation for the matrix model in the
large $N$ limit becomes
%(2.26)
\begin{equation}
 \biggl ( N^2 \frac{\partial}{\partial N^2}-\beta (g)\frac{\partial}
{\partial g}+1 \biggr ) y = \frac{2}{1+24g}.
\end{equation}

The exponent is obtained by
%(2.27)
\begin{equation}
\biggl ( \frac{d\beta (g)}{dg} \biggr ) _{g=g_c} = \frac{2}{3}
\end{equation}
\noindent
and
%(2.28)
\begin{equation}
  1 - \gamma_{st} = 1/\beta^{\prime} (g) \left|_{g=g_c} \right.
  = \frac{3}{2}.
\end{equation}
\noindent
The renormalization group equation for the free energy is obtained
from (2.23).  We define the $\beta$-function for the free energy
by $\tilde{\beta}(g)$,
%
%(2.29)
\begin{equation}
F(g) = \tilde{\beta}(g)\frac{\partial F}{\partial g}+ r(g)
\end{equation}
\noindent
Taking the derivative of (2.28), we have
%(2.30)
\begin{equation}
y = \tilde{\beta}(g)\frac{\partial y}{\partial g}
+ y\frac{\partial \tilde{\beta}(g)}{\partial g}
+ \frac{\partial r}{\partial g}.
\end{equation}
\noindent
Thus we are able to indentify
%
%(2.31)
\begin{equation}
\frac{\tilde{\beta}(g)}{1-\frac{\partial\tilde{\beta}(g)}{\partial g}}
=\beta (g) = -\frac{g}{3}
\biggl (\frac{1+48g}{1+24g} \biggr )
\end{equation}
%
%(2.32)
\begin{equation}
\biggl ( \frac{\partial r}{\partial g} {\biggr )}
\frac{1}{1-\frac{\partial\tilde{\beta}(g)}{\partial g}}
= \frac{2}{1+24g}.
\end{equation}
\noindent
Solving (2.31), we obtain $\tilde{\beta}$ as ,
%(2.33)
\begin{equation}
\tilde{\beta}(g) =-\frac{1}{2}g -\frac{36g^2}{1+48g}
+ \frac{864g^3}{(1+48g)^{3/2}}\ln
\left| \frac{1-\sqrt{1+48g} }{1+\sqrt{1+48g} } \right| .
\end{equation}

The expression for the $\beta$-function of this matrix model is very
similar to $O(N)$ vector model.  We find the following result by
the same analysis for the ratio of the coefficients,
%
%(2.34)
\begin{equation}
\frac{\partial F}{\partial g} = -\frac{1}{2}g
\frac{(1+32g)}{(1+24g)}\biggl ( \frac{\partial^2 F}{\partial g^2}
\biggr ) + \frac{2}{1+24g}
\end{equation}
\noindent
for the $O(N)$ vector model with the following Hamiltonian
%
%(2.35)
\begin{equation}
F = \frac{1}{2}r^2 + \frac{g}{N}(r^2 )^2
\end{equation}
\noindent
where $r^2 = \phi_1^2 + \cdots + \phi_N^2$.  The renormalization
group equation (2.33) is consistent with the coupled equations (2.15)
and (2.12).  We will discuss further the renormalization
group $\beta$-function in \S 8.

We apply our renormalized expansion method to the case of the
negative coefficient of $trM^2$.  The Hamiltonian is
%
%(2.36)
\begin{equation}
H = \frac{\alpha}{2}{\rm tr}M^2 + \frac{g}{N}{\rm tr}M^4
\end{equation}
\noindent
where we consider the region $\alpha < 0$, $g>0$, and $M$ is
$N \times N$ Hermitian matrix.  The effective Hamiltonian for this
model is the same as (2.7) except the coefficient of $x$ which now
becomes $\alpha /2$ instead of $1/2$.  The saddle point equation
(2.12) becomes
%(2.37)
\begin{equation}
\alpha x = 1 - 4gy
\end{equation}
\noindent
and the equations (2.13),(2.14) and (2.15) remain same.

There is a critical point, and beyond this point the saddle point
value $x$ is freezed.  This transition is common in various models
in the large $N$-limit.$^{11),12)}$  It is expected that $x$ is freezed as
%(2.38)
\begin{equation}
x = s \frac{\alpha}{g}
\end{equation}
\noindent
where $s$ is negative constant to be determined.
The critical point $x_c$ and $c$
are determined by the equations (2.35) and (2.15).  Inserting
$x=\frac{s\alpha}{g}$ into (2.15),
%(2.39)
\begin{equation}
\frac{s^2 \alpha^2}{g}= \frac{1}{108}
\biggl [ (4-\frac{3\alpha^2 s}{g})^{\frac{3}{2}} -1 \biggr ]
+ \frac{1}{12} - \frac{\alpha^2 s}{12g} .
\end{equation}
\noindent
Denoting $-\frac{s\alpha^2}{g}$ by $z$, we have
%(2.40)
\begin{equation}
-sz = \frac{1}{108} \biggl [ (4+3z)^{\frac{3}{2}}-1 \biggr ]
    + \frac{1}{12} + \frac{z}{12} .
\end{equation}
\noindent
The righthand side of above equation is represented in Fig.3.
The critical value at which the degenerate solution is obtained, is
$z=4$, $s=- {{1}\over{4}}$.  Then we find the critical point
%(2.41)
\begin{equation}
     \frac{\alpha^2}{g} = 16.
\end{equation}
\noindent
When $\frac{\alpha^2}{g}\geq 16$, the saddle point value of $x$ is
freezed as

%(2.42)
\begin{equation}
     x = - \frac{\alpha}{4g}.
\end{equation}
\noindent
{}From (2.37), we have for this case
%(2.43)
\begin{equation}
y = \frac{1-x}{4g} = \frac{1}{4g} + \frac{\alpha}{16g^2}
\end{equation}

%(2.44)
\begin{equation}
F = \frac{1}{4} \ln g - \frac{\alpha}{16g}.
\end{equation}

\noindent
This solution has been obtained before by the integral equation of the
eigenvalue.$^{10)}$  We have obtained this solution by the saddle point
method.

%section 3
\sect{Numerical estimation of the string susceptibility for
one matrix model}

\par

  Although one matrix model is exactly solvable, we consider the
numerical method to obtain the string susceptibility from the
renormalized perturbation.

The series for $y =(\frac{\partial F}{\partial g})$, the derivative
of the free energy is given by (2.13).
%(3.1)
\begin{equation}
gy = \sum _{K=1}^\infty c_K (gx^2 )^K .
\end{equation}
\noindent
This series is obtained from the irreducible diagram expansion.
This expansion becomes

%(3.2)
\begin{equation}
gy = 2t - 4t^2 + 32t^3 - 384t^4 + 5632t^5 - 93184t^6 + \cdots
\end{equation}

The ratio method gives
$t_c =gx^2 = - \frac{1}{48} \times (\frac{4}{3})^2 =-\frac{1}{27}$.
Denoting the ratio of the coefficients $c_K$ by $R_K$,
$R_K =c_K /c_{K-1}$,
%(3.3)
\begin{equation}
R_K \sim (-27)\biggl ( 1+\frac{-2+\gamma_{st}}{K} \biggr ) ,\;\; K\gg 1.
\end{equation}

{}From Fig.4, we obtain $\gamma_{st}=-\frac{1}{2}$.  Although this
analysis is worse than the ratio method for the free energy (2.18)
and (2.19), where the ratio is exactly given by a simple Pad\'e form,
it gives the correct value of $\gamma_{st}$.

%Fig.4
%Fig.5

As another numerical method for the perturbational series, we
investigate the saddle point equation (2.10).  Instead of the
infinite series, if we use the approximate finite series up to order
$(gx^2 )^K$, we find the critical value of $g$, beyond which there
is no real solution for the equation.  This analysis has been
investigated for the eigenvalue representation.$^{13)}$

The singularity of the finite order is the same as a branched
polymer case which has a following saddle point equation;
%(3.4)
\begin{equation}
     x + 8gx^2 - 1 = 0.
\end{equation}

\noindent
Thus the string susceptibility is $1/2$.  However increasing the
order of the approximation, the critical value $g_c$ approaches
to the value of $-1/48$, and the difference is scaled as

%(3.5)
\begin{equation}
48 - A_c (K) \sim \frac{1}{K^{1/\nu}}
\end{equation}

\noindent
where $A_c (K)=-1/g_c$ evaluated by the saddle point equation

%(3.6)
\begin{equation}
x + 8gx^2 - 16g^2 x^4 + 128g^3 x^6 + \cdots + c_K g^K x^{2K}-1=0.
\end{equation}

\noindent
The exponent $\nu$ is related to the string susceptibility

%(3.7)
\begin{equation}
     \gamma_{st} + 2\nu = 2.
\end{equation}

\noindent
For $\gamma_{st}=-\frac{1}{2}$, we have $\nu =\frac{5}{4}$.

The ratio behaves
%(3.8)
\begin{equation}
R_K = \frac{48-A_c (K)}{48-A_c (K-1)} \sim 1-\frac{1}{\nu K}.
\end{equation}

\noindent
The critical value of $A_c (K)$ is shown in Table 1.

%table 1

%section 4
\sect{$d=1$ one matrix model}

\par
  $d=1$ one matrix model is solved by the fermion formulation for the
inverted double well potential in the eigenvalue
representation.$^{1)}$  We consider this model by
the vector model representation.  The equivalent model is
%(4.1)
\begin{equation}
H = \frac{1}{2}(\nabla\vec{\phi})^2 + \frac{1}{2}\vec{\phi}^2
  + \sum_{K=1}^{\infty} \frac{c_K (\vec{\phi}^2 )^{2K}}{N^{4K-2}}g^K
\end{equation}

\noindent
where $\vec{\phi}$ is the $N^2$-dimensional vector field and the free
energy $F$ in the large $N$ limit is supposed to be identical of
$d=1$ one matrix model.  In one dimension, the problem becomes
a quantum mechanical one and the free energy is equal to the ground
state energy $\epsilon_0$ for the Schr\"{o}dinger equation,

%(4.2)
\begin{equation}
\biggl ( -\frac{1}{2}\frac{d^2}{dr^2}-\frac{N^2 -1}{2r}\frac{d}{dr}
+ \frac{1}{2}r^2 + \sum \frac{c_K g^K}{N^{4K-2}}(r^2 )^{2K} \biggr )
\psi = \epsilon_0 \psi .
\end{equation}

\noindent
The $N^2$-dimensional Laplacian is represented by the radial
coordinate.  To eliminate the first derivative term, the wave
function is written by

%(4.3)
\begin{equation}
\psi = r^{\alpha} \rho (r)
\end{equation}

\noindent
and $\alpha = (1-N^2 )/2$,
%
%(4.4)
\begin{equation}
\biggl( -\frac{1}{2}\frac{d^2}{dr^2}+ \frac{(N^2 -1)(N^2 -3)}{8r^2}
+ \frac{r^2}{2}+\sum_{K=1}^{\infty}
\frac{c_K g^K}{N^{4K-2}}(r^2 )^{2K} \biggr ) \rho
=\epsilon_0\rho .
\end{equation}

\noindent
Replacing $r$ by $Nr$, we have in the large $N$ limit,

%(4.5)
\begin{equation}
\biggl ( -\frac{1}{2}\frac{d^2}{dr^2}+N^2 V(r)\biggr ) \rho =
\epsilon_0\rho
\end{equation}

%(4.6)
\begin{equation}
\epsilon_0 = \frac{1}{8r^2}+\frac{r^2}{2}
           + \sum_{K=1}^{\infty}c_K g^K (r^2 )^{2K}
\end{equation}

\noindent
and $r$ is determined by

%(4.7)
\begin{equation}
\frac{dV(r)}{dr} = 0.
\end{equation}

\noindent
Using the notation $x=r^2$ as before, we have

%(4.8)
\begin{equation}
V(x) = \frac{1}{8x} +\frac{x}{2}+\sum_{K=1}^{\infty}c_K g^K x^{2K}
\end{equation}

\noindent
and the saddle point equation becomes from (4.7),

%(4.9)
\begin{equation}
-\frac{1}{8x^2}+\frac{1}{2}+\sum_{K=1}^{\infty}c_K g^K (2K)x^{2K-1}=0.
\end{equation}

\noindent
Denoting the derivative of $F$ for $g$ by $y$, we have a set of
equations, which are similar to (2.12) and (2.13),

%(4.10)
\begin{equation}
y = ( \frac{\partial F}{\partial g})
  = \sum_{K=1}^{\infty}c_K Kg^{K-1}x^{2K}
\end{equation}

%(4.11)
\begin{equation}
x = \frac{1}{4x} - 4gy.
\end{equation}

\noindent
{}From the direct perturbational series of $F$,$^{1),4)}$

%(4.12)
\begin{eqnarray}
F & = & \frac{1}{2} + 2g - 17g^2 + 300g^3 - 7126g^4 + 200112g^5  \nonumber \\
  & - & 6264798g^6 + 211756424g^7 - 7578384770g^8 + \cdots .
\end{eqnarray}

\noindent
{}From (4.11), we find

%(4.13)
\begin{eqnarray}
x & = & \frac{1}{2} - 4g + 84g^2 - 2344g^3 + 75776g^4 - 2684576g^5  \nonumber
\\
  & + & 101283112g^6 - 4001101488g^7 + 163717755104g^8
  + \cdots
\end{eqnarray}

\noindent
and we represent $y$ in the series of $x$ as

%(4.14)
\begin{eqnarray}
y &=& 8x^2 - 32gx^4 + 2304g^2x^6 - 137216g^3x^8 + 9093120g^4x^{10} \nonumber \\
  &-& 170676\times 2^{12}g^5 x^{12}
      + 3552248\times 2^{14}g^6 x^{14} \nonumber \\
  &-& 77649424\times 2^{16}g^7 x^{16}
  + \cdots .
\end{eqnarray}

\noindent
This equation determines the coefficient $c_K$ in (4.1).  We find
$c_1 =8$, $c_2 =-16$, $c_3 =768$, $c_4 =-34304$.

The saddle equation is solved by the successive approximation
of taking up to order $g^K$, and the critical value $g_c$ is
obtained as a point beyond which the root $x$ becomes complex.
In Table 2, these values of $A_c =-1/g_c$ are represented for
the order $g^K$.  The exact value of $A_c$ is known as

%(4.15)
\begin{eqnarray}
A_c (\infty ) & = & 12 \sqrt{2} \pi  \nonumber  \\
              & = & 53.3146 .
\end{eqnarray}

\noindent
In Table 2, the successive approximation approaches to this value.
To obtain the exponent $\nu$, we use the scaling behavior of the
finite order $K$,

%(4.16)
\begin{equation}
A_c (\infty ) = A_c (K) + \frac{c}{K^{1/\nu}}.
\end{equation}

\noindent
The ratio of the shift of cosmological constant $A_c$ becomes

%(4.17)
\begin{equation}
R_K = \frac{A_c (\infty )-A_c (K)}{A_c (\infty ) -A_c (K-1)}
    \sim 1 - \frac{1}{\nu K}.
\end{equation}

\noindent
In Fig.5, $R_K$ is plotted for $1/K$ and from the slope, we find
$\nu =1$.  This leads to the value of $\gamma_{st}$ by the
scaling relation $\gamma_{st}+2\nu =2$,

%(4.18)
\begin{equation}
\gamma_{st} = 0.
\end{equation}

%section 5
\sect{ Gauged matrix model}

\par
   In this section, we investigate a gauged matrix model.  The gauge
field is Abelian and the model is related to the Ginzburg-Landau
model in a strong magnetic field, which appears in the superconductor
with a magnetic field.$^{14)}$  We consider $d=2$ case.  In a strong
magnetic field, the lowest Landau level(L.L.L.) becomes important
since the energy level is quantized and the gap $\hbar \omega_c$
between the ground state and the first Landau level becomes large
for strong magnetic field.

  The gauged matrix model has been proposed for $d=2$ to avoid the
tachyonic instability.$^{15)}$  We show that our L.L.L. model is
well suited to the renormalized expansion formulation.

  As extensively studied for Ginzburg-Landau model, the free energy
is easily expanded by the coupling constant $g$.  Since the
Hilbert space is restricted to L.L.L., the order parameter is
expanded in the L.L.L. wave functions.  The degrees of two
dimensional coordinates are quantized and the system becomes zero
dimensional with the non-local interaction, which takes the
Gaussian form.  Usual L.L.L. model is written by a complex field.
We generalized this complex field $\phi$ to a complex matrix
$\phi_{ij}$.  The Hamiltonian of this system is given by
%(5.1)
\begin{equation}
H(\tilde{\phi}) = \frac{1}{2m}{\rm tr}|(-i\nabla_\mu -eA_\mu )
\tilde{\phi}|^2 + \alpha{\rm tr}|\tilde{\phi}|^2
+ \frac{\beta}{2N} {\rm tr}|\tilde{\phi}|^4
\end{equation}
where $\tilde{\phi}$ is a rand $N$ complex matrix, and
$\mu =x,y$.  Subjecting on the lowest Landau level in a strong
magnetic field with a gauge choice $A=(0,xB,0)$, the order
parameter is expanded by a harmonic oscillator in $x$ coordinate,
%(5.2)
\begin{equation}
\tilde{\phi}_{ij}
= \sum_q M_{i,j}(q)(L_y )^{-\frac{1}{2}}
  (\frac{eB}{\pi})^{\frac{1}{4}}e^{iqy}
  \exp [-\frac{eB}{2}(x-\frac{q}{eB})^2 ],
\end{equation}
\noindent
where $M_{i,j}(q)$ is a rank $N$ complex matrix and $L_y$ is the
length of the system.  Then Hamiltonian becomes
%(5.3)
\begin{eqnarray}
H & = & \alpha_B \sum_q {\rm tr}|M(q)|^2
       + \sum_{q_i}\frac{\beta}{2L_y N}(\frac{eB}{2\pi})^{\frac{1}{2}}
       \exp [-\frac{1}{2eB}
       (\sum q_i^2 -\frac{1}{4}(\sum q_i )^2 )] \nonumber \\
  & \times & \delta_{q_1 +q_2 ,q_3 +q_4}{\rm tr}
       [M(q_1 )^* M(q_3 )M(q_2 )^* M(q_4 )],
\end{eqnarray}
\noindent
where $\alpha_B =\alpha +(e/4M)B$.  We denote
$eB\beta /4\pi\alpha_B^2$ by $g$.  The nonlocal interaction appears
but the form of this interaction is Gaussian in one dimensional
momentum coordinate.  The perturbation for the free energy reduces
to the Gaussian integral of $q$ variables.  This integral is
equivalent to counting the number $T$ of Euler path on each
Feynman diagram.  The each diagram has $1/T$ extra factor to the
usual one matrix combinatorial factor.

  We consider the planar diagrams.  The Free energy is expanded
up to order $g^8$ as$^{16)}$
%(5.4)
\begin{eqnarray}
F & = &
     2g-17g^2 +248.8888g^3 -4687.4666g^4 +102344.29g^5 \nonumber \\
  & - &
     2464055.7g^6 +63603713.1g^7 -1729741366.0g^8 .
\end{eqnarray}

\indent
  This perturbation expansion has been evaluated for the
unrenormalized scheme.  We now discuss this model by the
renormalized expansion scheme as shown in \S 2.

  We consider the equivalent $N^2$-vector model, which has the
following free energy
%(5.5)
\begin{equation}
F = \frac{1}{2}x + \sum_{K=1}^{\infty}c_K g^K x^{2K}
    - \frac{1}{2}\ln x.
\end{equation}
\noindent
The number of Euler path $T$ becomes for the diagrams in Fig.2 as
(a) $T=1$, (b) $T=2$, (c) $T=3$, (d) $T=4$, (e) $T=5$, (f) $T=5$,
(g) $T=7$, (h) $T=8$.  Then together with the combinatoric factors
shown in Fig.2, we have
%(5.6)
\begin{equation}
F=-\frac{1}{2}\ln x +\frac{1}{2}x+2gx^2 -g^2 x^4 +\frac{32}{9}g^3 x^6
 - 20.8g^4 x^8 +\frac{441344}{2800}g^5 x^{10} - \cdots .
\end{equation}
\noindent
The saddle point equation is given by
%(5.7)
\begin{eqnarray}
x(\frac{\partial F}{\partial x})
& = &
      \frac{1}{2}x-\frac{1}{2}+4gx^2 -4g^2x^4 + \frac{64}{3}g^3 x^6
       - \cdots  \nonumber \\
& = & 0.
\end{eqnarray}
\noindent
This equation is written as
%(5.8)
\begin{equation}
x = 1 - 4gy
\end{equation}
\noindent
where $y=(\partial F/\partial g)$.

  This renormalized scheme of (5.6) and (5.7) becomes possible
since the factor $T$ of the number of Euler path is factorized.
This is different from the nonlocal propagator considered in
\S 4.

  The ratio method for the singularity of the free energy gives
$\gamma_{st}=0$.  Contrary to the $d=1$ matrix model,$^{2)}$ it seems
that there is no logarithmic singularity $\ln K$ term in the ratio
$R_K =c_K /c_{K-1}$, where $c_K$ is the coefficient of the free
energy of order $g^K$ in (5.4)
%(5.9)
\begin{equation}
R_K = A(1 + \frac{-3+\gamma_{st}}{K}).
\end{equation}
\noindent
We note that Penner model$^{17)}$ has also no $\ln K$ term
in the ratio, which shows
$\gamma_{st}=0$.

  We consider that this gauged model belongs to the universality
class of $2D$ gravity coupled to $c=1$ matter field, although other
gauged matrix model belongs to $c=2$ case.$^{14)}$

  We solve the saddle point equation by the approximation of order
$(gx^2 )^K$ and find the critical point $g_c$.
These values are approaching
to $A_c \cong 40$.  As same as the analysis of section 2, we deduce the
critical exponent $\nu$ by the ratio method, and find $\nu =1$,
$\gamma_{st}=0$.  In \S 8, we will discuss the renormalization group
function for this model.
%
%section 6
\sect{Two-matrix model}

\par
  Two-matrix model has the following Hamiltonian,
%(6.1)
\begin{equation}
H = \frac{1}{2}{\rm tr}M_1^2 +\frac{1}{2}{\rm tr}M_2^2
  + a{\rm tr}M_1 M_2 +\frac{g}{N}({\rm tr}M_1^4 +{\rm tr}M_2^4 )
\end{equation}
\noindent
where $M_1$ and $M_2$ are rank $N$ Hermitian matrices.  The
parameter $a$ is a coupling constant between two-matrices.  This
model represents the Ising model on a random surface,$^{18)}$
and the parameter $a$ is related to the Ising interaction $\beta$
as $a=e^{-2\beta}$.
By the unitary matrix, this Hamiltonian is diagonalized and the
large $N$ limit can be solved.$^{19)}$

  The perturbational series for this model becomes$^{4)}$
%(6.2)
\begin{eqnarray}
\frac{F}{2}
& = & 2g-g^2 \{ 16(1+a^2 )+2(1+a^4 )\}
      + g^3\{ 128(1+a^2 )^2 +\frac{256}{3}(1+3a^2 ) \nonumber \\
& + & \frac{32}{3}(1+3a^4 )+64(1+a^2 +2a^4 ) \}+O(g^4 )
\end{eqnarray}

\noindent
where the polynomials of $a^2$ appear for each diagrams shown in
Fig.1.  There are two vertices $M_1$ and $M_2$, and a factor $a$
appears for each bond which connects the different colour vertices.
The vertices $M_1$ and $M_2$ correspond to the spin-up and
spin-down of the Ising spins which are placed on the vertices.

  This expansion is based on the unrenormalized expansion as shown
in Fig.1.  It may be useful to find the renormalized expansion
scheme as one matrix model or a gauged model for writing the model
as $N^2$-vecotr model.  Since the factor of the polynomials of $a$
is somehow nonlocal and it is not factorized in a simple way.
We diagonalize the Hamiltonian of $M_1$ and $M_2$ by introducing
$L_1$, $L_2$ as
%(6.3)
\begin{eqnarray}
L_1 & = & (M_1 +M_2 )/\sqrt{2} \nonumber \\
L_2 & = & (M_1 -M_2 )/\sqrt{2} .
\end{eqnarray}

We have by this transformation,
%(6.4)
\begin{eqnarray}
H &=&
  \frac{1}{2}{\rm tr}(L_1^2 +L_2^2 )+\frac{g}{2(1-a^2 )^2 N}{\rm tr}
  \{ (1-a)^2 L_1^4 +(1+a)^2 L_2^4 \nonumber \\
  &+&
  4(1-a^2 )L_1^2 L_2^2 +2(1-a^2 )
  L_1 L_2 L_1 L_2 \} .
\end{eqnarray}

\noindent
Replacing $L_1$ and $L_2$ by $L_1 /\sqrt{(1-a)}$ and
$L_2/\sqrt{(1+a)}$, and $g/2(1-a^2 )^2$ by $g$, we have
%(6.5)
\begin{eqnarray}
F &=& \frac{x_1}{2(1-a)}+\frac{x_2}{2(1+a)}-\frac{1}{2}\ln x_1 \nonumber \\
  &-& \frac{1}{2}\ln x_2 + 2g[x_1^2 +x_2^2 + 2x_1 x_2 ]
     - 2g^2 [x_1^4 +x_2^4 +6x_1^2 x_2^2 ] \nonumber \\
  &+& \frac{32}{3}g^3 [x_1^6 +x_2^6 + 3x_1^4 x_2^2 +3x_1^2 x_2^4
     + 8x_1^3 x_2^3 ]+ 0(g^4 )
\end{eqnarray}

\noindent
where we used two variables $x_1$ and $x_2$ for ${\rm tr}L_1^2$
and ${\rm tr}L_2^2$.  The saddle point conditions for
$x_1$ and $x_2$ become,
%(6.6)
\begin{equation}
x_1\frac{\partial F}{\partial x_1} =0, \;\;\;\;
x_2\frac{\partial F}{\partial x_2}=0
\end{equation}
\noindent
which lead to
%(6.7)
\begin{equation}
\frac{x_1}{2(1-a)}-\frac{1}{2}+2g[2x_1^2 +2x_1 x_2]
 -2g^2 [4x_1^4 +12x_1^2 x_2^2 ] +0(g^3 )  =  0,
\end{equation}
%(6.8)
\begin{equation}
\frac{x_2}{2(1+a)}-\frac{1}{2}+2g[2x_2^2 +2x_1 x_2]
 -2g^2 [4x_2^4 +12x_1^2 x_2^2 ] +0(g^3 )  =  0,
\end{equation}

\noindent
we find the following identity, corresponds to (2.12)
%(6.9)
\begin{equation}
\frac{x_1}{2(1-a)}+\frac{x_2}{2(1+a)}
= 1 - 2g \biggl ( \frac{\partial F}{\partial g} \biggr ).
\end{equation}

   Using the saddle point equations (6.7) and (6.8), we obtain
the series expansion for the free energy $F$ as
%(6.10)
\begin{equation}
F = 1 + 8g -g^2 (128(1+a^2 )+16(1+a^4 ))+ \cdots
\end{equation}

\noindent
which becomes the same result of ref.4) when $g$ is replaced by
$g/2$ and $F$ is devided by a factor 2.
Two matrix model has a Ising phase transition at $a=1/4$.$^{18)}$
First, we try to solve the saddle point equations in the
successive approximation, which takes acount the terms up to order
$g^K$.  The coupled saddle point equation is easily solved
numerically for example by plotting two solutions
of (6.7) and (6.8) in $x_1 -x_2$ plane.  The degenerate solution
point gives $g_c$.  The obtained $g_c$ is shown in Table 3.

  The exact value of $A_c =-1/g_c$ is 101.25.  The shift of the
critical point for the finite order approximation is analyzed by
the same ratio method as (3.7),
%(6.11)
\begin{equation}
R_K = \frac{A_\infty -A_K}{A_\infty -A_{K-1}}=1 -\frac{1}{\nu K}.
\end{equation}

\noindent
The obtained value of $\nu$ is consistent with the exact value of
%(6.12)
\begin{equation}
\nu = 1-\frac{\gamma_{st}}{2} = \frac{7}{6}.
\end{equation}

\noindent
The string susceptibility $\gamma_{st}$ for the two matrix model is
$\gamma_{st}=-1/3$.

  Our renormalized expansion gives the series expansion for
$y=(\partial F/\partial g)$, as a function of $x_1$ and $x_2$.
We parametrize
%(6.13)
\begin{equation}
x_1 = \lambda x_2 .
\end{equation}
\noindent
Then we have
%(6.14)
\begin{equation}
gy = \sum_{K=1}^{\infty}c_K (gx_2^2 )^K
\end{equation}

\noindent
with
%(6.15)
\begin{eqnarray}
c_1 &=& 2(1+\lambda )^2 \nonumber \\
c_2 &=& -4(1+6\lambda^2 +\lambda^4) \nonumber \\
c_3 &=& 32(1+3\lambda^2 +8\lambda^3+3\lambda^4+\lambda^6) \nonumber \\
c_4 &=& -384(1+\frac{8}{3}\lambda^2+\frac{16}{3}\lambda^3
      + 14\lambda^4+\frac{16}{3}\lambda^5+\frac{8}{3}\lambda^6
      +\lambda^8) \nonumber \\
c_5 &=& 5632(1+\frac{30}{11}\lambda^2+\frac{50}{11}\lambda^3
      +\frac{135}{11}\lambda^4+\frac{252}{11}\lambda^5 \nonumber \\
    &+& \frac{135}{11}\lambda^6+\frac{50}{11}\lambda^7
      +\frac{30}{11}\lambda^8+\lambda^{10}) \nonumber \\
c_6 &=& -93184(1+\frac{264}{91}\lambda^2
      +\frac{448}{91}\lambda^3 +\frac{1029}{91}\lambda^4
      +\frac{2156}{91}\lambda^5 \nonumber \\
    &+& \frac{3672}{91}\lambda^6 +\frac{2156}{91}\lambda^7
      +\frac{1029}{91}\lambda^8 +\frac{448}{91}\lambda^9
      +\frac{264}{91}\lambda^{10}+\lambda^{12}).
\end{eqnarray}

\noindent
These polynomials of $\lambda$ are easily obtained from the
diagram representation in Fig.2.  The coefficients are the
combination number of the $x_1$-lines.  At each vertex, the number
of $x_1$ line should be either 0, 2 or 4.

  Although we could obtain the string susceptibility from (6.14) at
$\lambda =0$ as (3.3) for the one matrix model, we have no
information about the crossover at the critical value
$\lambda_c$ from the series (6.14).

  It may be necessary to reconstruct the series expansion about
the coupling constant $g$ for the string susceptibility as shown
in ref. 4).  We will discuss the renormalization group function
for this model in \S 8.

%section 7
\sect{$n$-Ising model on a random surface}

\par
  On each vertex of the Feynman diagram, different $n$-species
Ising spins are placed.  This model represents 2d quantum gravity
coupled to matter field of the central charge $c=n/2$, since one
Ising plays the role of $c=1/2$.  It is easily represented by the
$2^n$ matrix model.  For example, $n=1$ case is the previous
two-matrix model and in the $n=2$ case we have the following
four-matrix model,
%(7.1)
\begin{eqnarray}
H & = & \frac{1}{\rm tr}(M_1^2 +M_2^2 +M_3^2 +M_4^2 )
      + a{\rm tr}(M_1M_2 +M_1M_3 +M_2M_4 +M_3M_4 ) \nonumber \\
  & + & a^2{\rm tr}(M_1M_4 +M_2M_3 )
      +\frac{g}{N}{\rm tr}(M_1^4 +M_2^4 +M_3^4 +M_4^4 )
\end{eqnarray}

\noindent
This Hamiltonian is diagonalized by
%(7.2)
\begin{equation}
\left (
\begin{array}{c}
M_1 \\
M_2 \\
M_3 \\
M_4
\end{array}
\right )
=
\frac{1}{\sqrt{2}}
\left (
\begin{array}{crrr}
1  &  1  & -1  & -1 \\
1  & -1  &  1  & -1 \\
1  &  1  &  1  &  1 \\
1  & -1  & -1  &  1
\end{array}
\right )
\left (
\begin{array}{c}
L_1 \\
L_2 \\
L_3 \\
L_4
\end{array}
\right )
\end{equation}

\noindent
and we have
%(7.3)
\begin{equation}
H_0 =
    \frac{1}{2}\sum_{i=1}^4{\rm tr}M_i^2
    = (1+a)^2\frac{L_1^2}{2}+(1-a^2 )\frac{L_2^2}{2}
    + (1-a)^2\frac{L_3^2}{2}+(1-a^2 )\frac{L_4^2}{2}.
\end{equation}

\noindent
Using the replacement $L_1\rightarrow L_1/(1+a)$,
$L_2 \rightarrow L_2/\sqrt{1-a^2}$,
$L_3 \rightarrow L_3/(1-a)$ and $L_4 \rightarrow L_4/\sqrt{1-a^2}$,
$g \rightarrow g/(1-a^2 )^4$
we have
%(7.4)
\begin{eqnarray}
H & = & {\rm tr}\{ \frac{L_1^2}{2(1-a)^2}+\frac{L_2^2}{2(1-a^2)}
     +\frac{L_3^2}{2(1+a)^2}+\frac{L_4^2}{2(1-a^2 )} \} \nonumber \\
  & + & \frac{g}{N}{\rm tr}\{ L_1^4 +L_2^4 +L_3^4 +L_4^4 \nonumber \\
  & + & 4(L_3^2 L_4^2 +L_1^2 L_3^2 +L_2^2L_3^2 +L_1^2L_4^2 +L_2^2L_4^2
     + L_1^2L_2^2 ) \nonumber \\
  & + & 2(L_3L_4L_3L_4 + L_1L_3L_1L_3 +\cdots + L_1L_2L_1L_2 )
     + 24L_1L_2L_3L_4 \} .
\end{eqnarray}

\noindent
Thus we have the expression for the free energy written by
$x_1,\cdots ,x_4$, which are defined by
%(7.5)
\begin{equation}
x_i = {\rm tr}L_i^2.
\end{equation}
%(7.6)
\begin{eqnarray}
F &=& \frac{x_1}{2(1-a)^2}+\frac{x_2}{2(1-a^2)}
    +\frac{x_3}{2(1+a)^2}+\frac{x_4}{2(1-a^2)}
    -\frac{1}{2}\sum_{i=1}^4 \ln x_i \nonumber \\
  &+& 2g( x_1^2 +x_2^2 +x_3^2 +x_4^2 +2\sum_{i<j}x_ix_j )+O(g^2 ).
\end{eqnarray}

\noindent
The saddle point equations become
%(7.7)
\begin{equation}
x_1 \frac{\partial F}{\partial x_i} = 0 \;\;\; (i=1,\cdots ,4).
\end{equation}

\noindent
Thus we get
%(7.8)
\begin{equation}
\frac{x_i}{2f_i (a)}-\frac{1}{2}+4gx_i^2 +4gx_i (\sum_{j\neq i}x_j )
    + \cdots =0,
\end{equation}

\noindent
where $f_1 (a)=(1-a)^2$, $f_2 (a)=(1-a^2 )$, $f_3 (a)=(1+a)^2$
and $f_4 (a)=f_2 (a)$.  We find the following identity,
%(7.9)
\begin{equation}
\sum_{i=1}^4 \frac{x_i}{2f_i (a)}=2-2g\biggl ( \frac{\partial F}
    {\partial g} \biggr ) .
\end{equation}

\noindent
The terms of order $g^K$ in the expression for the free energy $F$
are essentially same as one matrix model.  The lines of Feynman
diagrams now have possibility to choose one of 4-colours denoted
by $x_i$.  The rule is that the colour of two lines at least should be
same at each vertices.  Thus, the polynomial of $x_i$ for each
irreducible diagram is determined.

  This rule is applied also to $n$-Ising model with $2^n$ matrix
representation.  The factors $f_i (a)$ for $2^N$-matrix model
$(i=1,\cdots ,2^n )$ are obtained by the diagonalization of the
Hamiltonian.  It is enough to calculate the unperturbed term of
the Hamiltonian $H_0$ for the determination of $f_i (a)$.
Using the notation $L_i$ of (7.2), we have
similar expression as (7.3).  For $n=3$, we get
%(7.10)
\begin{eqnarray}
H_0
 &=& \frac{(1+a)^3 L_1^2}{2}+\frac{(1+a)^2 (1-a)}{2}
      (L_2^2+L_3^2+L_4^2) \nonumber \\
 &+& \frac{(1+a)(1-a)^2}{2}(L_5^2+L_6^2+L_7^2 )
    + \frac{(1-a)^3}{2}L_8^2.
\end{eqnarray}

Denoting $g/(1-a^2)^{2n}$ by $g$, we have for $n=3$,
%(7.11)
\begin{eqnarray}
f_1(a)\! &=&\! (1-a)^3, f_2(a)=f_3(a)=f_4(a)=(1-a)^2(1+a) \nonumber \\
f_5(a)\! &=&\! f_6(a)=f_7(a)=(1+a)^2(1-a),f_8(a)=(1+a)^3.
\end{eqnarray}

\noindent
Thus we have general rules for writing the $2^n$-matrix model in the
vector representation.

  We have checked that the perturbational series for the free
energy $F$ obtained before$^{4)}$ for the $n$-Ising model is
recovered by this vector representation, in the replacement of
$g \rightarrow g/2$ as
%(7.12)
\begin{eqnarray}
\frac{F}{2^{n}}
 &=& \frac{1}{2}+2g-g^2 \{ 16(1+a^2)^n +2(1+a^4)^n \}
    +g^3\{ 128(1+a^2)^{2n}+\frac{256}{3}(1+3a^2)^n \nonumber \\
 &+& \frac{32}{3}(1+3a^4)^n+64(1+a^2 +2a^4)^n \}+O(g^4).
\end{eqnarray}

%section 8
\sect{Renormalization group function for the
matrix model}

\par
  As we have seen \S 2, the renormalization group function
$\beta (g)$ is simply given by (2.25) for the one matrix model.
This expression shows that $\beta$-function is regular at $g=0$,
and becomes zero at the negative coupling constant $g_c$.  We
may have the following  differential equation, which leads to
the renormalization group equation,
%(8.1)
\begin{equation}
\biggl ( \sum_{K=0}^\infty a_K g^K \biggr )
  \biggl (\frac{\partial F}{\partial g} \biggr )
= -g\biggl (\sum_{K=0}^\infty b_K g^K \biggr )
  \biggl (\frac{\partial^2 F}{\partial g^2} \biggr )
  +\biggl (\sum_{K=0}^\infty c_K g^K \biggr )
\end{equation}

\noindent
and the $\beta$-function becomes
%(8.2)
\begin{equation}
\beta (g)=-g \frac{\sum_{K=0}^\infty b_K g^K}{\sum_{K=0}^\infty a_K g^K}
\end{equation}

\noindent
where we put $a_0 =1$.

  We determine again the $\beta$-function of the one matrix model,
or equivalently the values of the coefficients $a_K$ and $b_K$
in (8.1).  As discussed before, the successive approximation
for the saddle point equation is made by taking the higher order
terms of the irreducible diagrams of the free energy.  The first
order approximation takes the following free energy,
%(8.3)
\begin{equation}
F = \frac{1}{2}x -\frac{1}{2}\ln x + 2gx^2.
\end{equation}

\noindent
The saddle point equation becomes
%(8.4)
\begin{equation}
\frac{1}{2}x -\frac{1}{2}+4gx^2 = 0.
\end{equation}

  The derivative $y=(\partial F/\partial g)$ is related to $x$
as (2.12), and we have in this order,
%(8.5)
\begin{equation}
y=2x^2 =2-32g +0(g^2 ).
\end{equation}

\noindent
In the large $g$ limit, we have $x \sim 1/\sqrt{8g}$ and
%(8.6)
\begin{equation}
y=\frac{1}{4g}-\frac{1}{8\sqrt{2}g^{3/2}}+(\frac{1}{g^2}).
\end{equation}

\noindent
{}From (8.5) and (8.6) we pose these asymptotic behavior conditions
on (8.1) as
%(8.7)
\begin{eqnarray}
2a_1 - 32 & = & 32b_0 \nonumber \\
     a_1  & = & b_1 + 8 \nonumber \\
      a_1 & = & \frac{3}{2}b_1 .
\end{eqnarray}

\noindent
Solving these coupled equations for the coefficients, we have
$a_1 =24$, $b_0 =\frac{1}{2}$ and $b_1 =16$, and the differential
equation is
%(8.8)
\begin{equation}
(1+24g)\biggl (\frac{\partial F}{\partial g}\biggr )
=
-\frac{1}{2}g(1+32g)\biggl (\frac{\partial^2 F}{\partial g^2}\biggr )
+ 2.
\end{equation}

\noindent
This equation is exact for the free energy of (8.3), which
corresponds to $N$-vector model (2.34).

  The next order $g^2$ is taken for the free energy as
%(8.9)
\begin{equation}
F=\frac{1}{2}x-\frac{1}{2}\ln x+2gx^2 -2g^2 x^4 .
\end{equation}

\noindent
The saddle point equation determines the value of $x$ up to order
$g^2$, and consequently determines the value of $y$ as
%(8.10)
\begin{equation}
y=\frac{1-x}{4g}=2-36g+0(g^2 )
\end{equation}

\noindent
The large $g$ limit also determines the value of $y$ as
%(8.11)
\begin{equation}
y \simeq \frac{1}{4g}-\frac{1}{8g^{3/2}}+O(\frac{1}{g^2}).
\end{equation}

\noindent
Thus we determine the values of $a_1$, $b_0$ and $b_1$, assuming
$a_K =b_K =0$ for $K\geq 2$,
%(8.12)
\begin{eqnarray}
2a_1 -36 & = & 36b_0 \nonumber \\
     a_1 & = & b_1 +8 \nonumber \\
     a_1 & = & \frac{3}{2}b_1 .
\end{eqnarray}

\noindent
The solution of above equation gives $a_1 =24$, $b_1 =16$ and
$b_0 =\frac{1}{3}$, and it leads to
%(8.13)
\begin{equation}
(1+24g)\biggl (\frac{\partial F}{\partial g}\biggr )
=
-\frac{1}{3}g(1+48g)\biggl (\frac{\partial^2 F}{\partial g^2}\biggr )+2
\end{equation}

\noindent
which becomes an exact equation of all orders of $g$.
For the one matrix model, the renormalization group $\beta$-function
(8.2) turns out to be simple.

  Next we consider the gauge model in \S 5.  Up to
order $g^2$, we have the saddle point equation,
%(8.14)
\begin{equation}
\frac{1}{2}x -\frac{1}{2}+4gx^2 -4g^2 x^4 =0.
\end{equation}

\noindent
{}From the results of the expansion of $y$ in the small $g$ and
the large $g$ behavior (5.7) and (5.8),
we determine $b_0 =\frac{7}{17}$,
$a_1 =24$ and $b_1 =\frac{272}{17}$ as
%(8.15)
\begin{equation}
(1+24g)\biggl ( \frac{\partial F}{\partial g}\biggr )
=
-\frac{7}{17}g\biggl ( 1+\frac{272}{7}g\biggr )
  \biggl ( \frac{\partial^2 F}{\partial g^2}\biggr ) +2.
\end{equation}

\noindent
This leads to
%(8.16)
\begin{equation}
\gamma_{st}=1-\frac{1}{\beta^{\prime}(g_c )}=\frac{1}{14}.
\end{equation}

  The next order of the approximation for the $\beta$-function is
%(8.17)
\begin{equation}
(1+a_1 g+a_2 g^2 )y=-g(b_0 +b_1 g+b_2 g^2 )(\frac{dy}{dg})
  + c_0 +c_1 g.
\end{equation}

\noindent
{}From the large $g$ behavior of $y$, we have
%(8.18)
\begin{equation}
y \simeq \frac{1}{4g} - \frac{\lambda}{g^{3/2}}
\end{equation}
%(8.19)
\begin{equation}
a_2 = 12c_1, \;\;\; b_2 = \frac{2}{3}a_2 .
\end{equation}

\noindent
The other coefficients in (8.17) are determined from the small $g$
expansion for $y$,
%(8.20)
\begin{equation}
y=2-34g +\frac{2240}{3}g^2 -\frac{281248}{15}g^3
  + \frac{53730752}{105}g^4 +O(g^5 ).
\end{equation}

\noindent
We have $a_1 =-33.528$, $a_2 =-1366.26$, $b_0 =0.37642$,
$b_1 =-8.3461$, $b_2 =-910.84$.  The $\beta$-function,
%(8.21)
\begin{equation}
\beta (g)
= -g\frac{b_0 +b_1 g+b_2 g^2}{1+a_1 g+a_2 g^2}
\end{equation}

\noindent
has a zero at $g_c=-1/39.3385$, and the derivative of $\beta (g)$
at $g_c$ becomes 0.995426.  Therefore we have
%(8.22)
\begin{equation}
\gamma_{st} =1-1/\beta^{\prime}(g)=-0.004595.
\end{equation}

\noindent
Although this $\beta$-function (8.21) has an unphysical singularity
in the positive $g$, it describes reasonably the behavior at the
critical point $-1/g_c =39.3385$ and the value of the derivative.
Thus we find that the string
susceptibility $\gamma_{st}$ is zero within our approximation.
This result is consistent with the result of $c=1$.

  We apply the form of $\beta$-function for the two-matrix model.
In the first simple approximation, we put
%(8.23)
\begin{equation}
(1+a_1 g)y = -g(b_0 +b_1 g)(\frac{dy}{dg})+c_0 ,
\end{equation}

\noindent
where $a_1$, $b_0$ and $b_1$ depend upon the coupling parameter
$a$.  At the critical value $a=\frac{1}{4}$, we obtain $a_1 =25.89$,
$b_0 =0.3623$, $b_1 =18.35$, $c_0 =2$.  The critical value $g_c$
is obtained as $-1/g_c =50.66$ while the exact value is
$-1/g_c =50.625$.  The estimated string susceptibility becomes
$\gamma_{st}=-0.3494$.  This approximation, however, does not give
the maximum peak at $a=1/4$.  Therefore, it is necessary to
evaluate higher order approximation.  The next order
approximation for the two-matrix model $\beta$-function
takes the form of (8.21) with $a_1 =34.941$, $a_2 =232.12$,
$b_0 =0.35932$, $b_1 =21.367$, $b_2 =161.45$, $c_0 =2$, and
$c_1 =18.206$.  The critical point becomes $-1/g_c =50.58$ and
$\gamma_{st}=-0.3501$ at $a=1/4$.  The form of our
differential equation is similar to the previous Pad\'e form by
the ratio method.$^{4)}$  And the result is  expected to be similar
to the previous analysis by the ratio method.$^{5),6)}$
The previous ratio method is restricted to [2,1] Pad\'e form.
Our differential form (8.1) is more general and precise in this
respect.

%section 9
\sect{Discussion}

\par
  In this paper, we have analyzed the matrix model by the
renormalized expansion method.  We have obtained equivalent
$N^2$-vector model which consists of the polynomial of
$({\rm tr}M^2 )^l$, and developed a new series expansion
by the help of the saddle point equation.
This method is effective for several matrix
models.  We have derived an exact renormalization group equation
for the one-matrix model and approximated one for the other matrix
models.  The approximation for the $\beta$-function is consistent
with the asymptotic behavior in the large coupling constant $g$
and also with the small $g$ expansion.  It has $[n,n]$ Pad\'e form,
and its derivative at the critical value $g_c$ gives the string
susceptibility.  This method is more systematic or precise
than the simple ratio method.$^{4)}$

  We have also investigated the shift of the critical point
in the successive approximation, which includes the irreducible
diagrams of order $g^K$.  The obtained string susceptibilities
due to the scaling relation are reasonable.  For $n$-Ising model
on a random surface, we obtained new perturbational expressions
which may be useful in the large $n$ case.  The large $n$ limit
is considered to be equivalent to the branched polymer
case.$^{20)}$  We will discuss the
large $n$, or large central charge $c$ case elsewhere.

\vspace{10mm}
\begin{center}
{\bf Acknowledgement}
\end{center}

  The author thanks E. Br\'ezin and S. Higuchi for the discussion
of the renormalization group method of the matrix model.  This work
is supported by the cooperative research project between Centre
national de la Recherche scientifique (CNRS) and Japan Society for
the Promotion of Science (JSPS).

\newpage

\begin{center}
{\bf References}
\end{center}
\vspace{3mm}

\begin{description}
\item[{1)}] E. Br\'ezin, C. Itzykson, G. Parisi and J.B. Zuber,
            Comm. Math. Phys. {\bf 59} (1978), 35.
\item[{2)}] M.R. Douglas and S.H. Shenker, Nucl. Phys. {\bf B335}
            (1990), 635.\\
            D.J. Gross and A.A. Migdal, Phys. Rev. Lett.
            {\bf 64} (1990), 127.\\
            E. Br\'ezin and V.A. Kazakov, Phys. Lett.
            {\bf B236} (1990), 144.
\item[{3)}] V.G. Knizhnik, A.M. Polyakov and A.B. Zamolodchikov,
            Mod. Phys. Lett. {\bf A3} (1988), 819.\\
            F. David, Mod. Phys. Lett. {\bf A3} (1988), 1651.\\
            J. Distler and H. Kawai, Nucl. Phys. {\bf B321} (1989), 509.
\item[{4)}] E. Br\'ezin and S. Hikami, Phys. Lett. {\bf 283} (1992),
            203.\\
            S. Hikami and E. Br\'ezin, Phys. Lett. {\bf 295} (1992),
            209.
\item[{5)}] S. Hikami, Phys. Lett. {\bf B305} (1993), 327.
\item[{6)}] M. Wexler, Nucl. Phys. {\bf 410} (1993), 377.
\item[{7)}] C.A. Baillie and D. Johnston, Phys. Lett. {\bf B286}
            (1992), 44.\\
            J. Ambjorn, B. Durhuus, T. Jo'nsson, G. Thorleifsson,
            Nucl. Phys. {\bf B398} (1993), 568.\\
            M. Bowick, M. Falcioni, G. Harris and E. Marinari,
            Phys. Lett. {\bf B322} (1994), 316.
\item[{8)}] E. Br\'ezin and J. Zinn-Justin, Phys. Lett. {\bf B288}
            (1992), 54.\\
            J.W. Carlson, Nucl. Phys. {\bf 248} (1984), 536.
\item[{9)}] S. Higuchi, C. Itoi and N. Sakai, Phys. Lett. {B312}
            (1993), 88.\\
            S. Higuchi, C. Itoi, S. Nishigaki and N. Sakai, Phys. Lett.
            {B318} (1993), 63.
\item[{10)}] Y. Shimamune, Phys. Lett. {\bf 108} (1982), 407.\\
             G.M. Cicuta, L. Molinari and E. Montaldi,
             J. Phys. {\bf A23} (1990), L421.
\item[{11)}] D.J. Gross and E. Witten, Phys. Rev. {\bf D21} (1980),
             446.
\item[{12)}] S. Hikami and T. Maskawa, Prog. Theor. Phys. {\bf 67}
             (1982), 1038.
\item[{13)}] S. Hikami, Physica {\bf A204} (1994), 290.
\item[{14)}] G.J. Ruggeri and D.J. Thouless, J. Phys. {\bf F6}
             (1976), 2063.\\
             E. Br\'ezin, A. Fujita and S. Hikami, Phys. Rev. Lett.
             {\bf 65} (1990), 1949.
\item[{15)}] K. Demeterfi and I.R. Klebanov, in {\sl Quantum Gravity}
             Nishinomiya-Yukawa Symposium, edited by K. Kikkawa and
             M. Ninomiya, World Scientific 1993, Singapore.
\item[{16)}] S. Hikami, in {\sl "Correlation effects in low
             dimensional electron systems"}, edited by A. Okiji
             (Taniguchi Symposium (1993)).  Springer Solid State
             Series, Springer, 1994.
\item[{17)}] J. Distler and C. Vafa, Mod. Phys. Lett. {\bf A6}
             (1991), 259.
\item[{18)}] D.V. Boulatov and V.A. Kazakov, Phys. Lett. {\bf B214}
             (1988), 581.
\item[{19)}] M.L. Mehta, Commun. Math. Phys. {\bf 49} (1981), 327.
\item[{20)}] M.G. Harris and J.F. Wheater, a preprint (hep-th.9404174).\\
             B. Durhuus, a preprint (hep-th.9402052).\\
             M. Wexler, a preprint (NBI-HE-94-28).
\end{description}

\newpage
\begin{description}
 \item[Table 1.] The inverse of the critical value $A=-1/g_c$ for one
                 matrix model\\

\vspace{6mm}
\begin{tabular}{|c|c|c|c|c|c|c|c|}
\hline
K & 1& 2& 3& 4& 5& 6& $\infty$ \\
\hline
$A_c (K)$ & 32& 37.92& 40.43& 41.844& 42.768& 43.44& 48 \\
\hline
\end{tabular}

\vspace{20mm}
 \item[Table 2.] The inverse of the critical value $A=-1/g_c$
                 for $d=1$ matrix model\\

\vspace{6mm}
\begin{tabular}{|c|c|c|c|c|c|c|c|c|}
\hline
K & 1& 2& 3& 4& 5& 6& 7& $\infty$ \\
\hline
$A_c (K)$ & $24\sqrt{3}$& 44.29& 46.61& 47.97& 48.83& 49.42& 49.88&
$12\sqrt{2}\pi$ \\
\hline
\end{tabular}

\vspace{20mm}
 \item[Table 3.] The inverse of the critical value $A=-1/g_c$ for two
                 matrix model at $a=1/4$ \\

\vspace{6mm}
\begin{tabular}{|c|c|c|c|c|c|c|}
\hline
K & 1& 2& 3& 4& 5& $\infty$ \\
\hline
$A_c(K)$ & 70.86& 82.84& 87.76& 90.44& 92.17& 101.25\\
\hline
\end{tabular}
\end{description}

\newpage
\noindent
{\bf Figure captions}
\vspace{10mm}
\begin{description}
 \item[Fig.1] Planar Feynman diagrams for the free energy of one matrix
              model up to order $g^3$.
 \item[Fig.2] (a) Planar irreducible Feynman diagrams for the free
              energy of one matrix model up to order $g^5$.  The
              coefficients for diagrams $a\sim h$ are $2,-2,\frac{32}{3}$,
              $-64,-32,102.4,512,512$ respectively.\\
              (b) Diagrams in order $g^6$. The coefficients for diagrams
              $a\sim i$ are $-341\frac{1}{3},-2048$,\\
              $-4096,-1024$,
              $-4096,
             -1024,-682\frac{2}{3},-2048,-170\frac{2}{3}$ respectively.
 \item[Fig.3] The solution of Eq.(2.40) is shown for the transition of
              one matrix model with negative $\alpha$. The dotted line
              represents $z/4$.
 \item[Fig.4] The ratio $R_K =c_K /c_{K-1}$ is plotted against $1/K$.
              The dotted line represents the asymptotic solution of
              (3.3) with $\gamma_{st}=-1/2$ for the one matrix model.
 \item[Fig.5] The ratio $R_K$ of the successive approach to the
              critical point for $d=1$ one matrix model, plotted
              against $1/(K+1)$. The solid line represents $\nu = 1$
              and $\gamma_{st} = 0$.
\end{description}
\end{document}